\documentclass[conference]{IEEEtran}
\IEEEoverridecommandlockouts
\pdfoutput=1
\usepackage[noadjust]{cite}
\usepackage{amsmath}
\usepackage{breqn}
\usepackage{array}
\usepackage{mdwmath}
\usepackage{amssymb}
\usepackage{algorithm}
\usepackage[noend]{algpseudocode}
\usepackage{eqparbox}
\usepackage{stfloats}
\usepackage{url}
\usepackage{tikz}
\usepackage{float}
\usepackage{pgfplots}
\graphicspath{{../pdf/}{../jpeg/}}
\DeclareGraphicsExtensions{.pdf,.jpeg,.png}
\usepackage{epstopdf}
\usepackage{amsfonts}
\usepackage{tikz}
\usepackage{ctable}

\usepackage{subfig}

\usepackage{stackengine}

\def\BibTeX{{\rm B\kern-.05em{\sc i\kern-.025em b}\kern-.08em T\kern-.1667em\lower.7ex\hbox{E}\kern-.125emX}}

\usepackage{tikz}
\usetikzlibrary{%
patterns,%
calc,%
fit,%
arrows,%
plotmarks,%
shadows,%
chains,%
shapes%
}

\usepackage{textcomp}

\begin{document}

\title{\huge Improving UAV Communication in Cell Free MIMO Using a Reconfigurable Intelligent Surface}   

\author{\IEEEauthorblockN{Bayan Al-Nahhas, Anas Chaaban, and Md. Jahangir Hossain}
\IEEEauthorblockA{School of Engineering, University of British Columbia, Kelowna, Canada. \\ 
Email:\{bayan.alnahhas, anas.chaaban, jahangir.hossain\}@ubc.ca}}

\maketitle

\begin{abstract}

Communication with unmanned aerial vehicles (UAVs) in current terrestrial networks suffers from poor signal strength due to the down-tilt of the access points (APs) that are optimized to serve ground users ends (GUEs). To solve this, one could tilt the AP antenna upwards or allocate more power to serve the UAV. However, this negatively affects GUE downlink (DL) rates. In this paper, we propose to solve this challenge using a reconfigurable intelligent surface (RIS) to enhance the UAV communication while preserving the 3GPP-prescribed downwards antenna tilt and potentially improving the DL performance of the GUE. We show that under conjugate beamforming (CB) precoding and proper power split between GUEs and the UAV at the APs, an RIS with phase-shifts configured to reflect radio signals towards the UAV can significantly improve the UAV DL throughput while simultaneously benefiting the GUEs. The presented numerical results show that the RIS-aided system can serve a UAV with a required data rate while improving the GUEs DL performance relative to that in a CF-MIMO system without a UAV and an RIS. We support this conclusion through simulations under a varying numbers of RIS reflecting elements, UAV heights, and power split factor.

\end{abstract}

\section{Introduction}

\label{Sec:Intro}

With the rapid development of fifth generation (5G) wireless networks, unmanned aerial vehicles (UAVs) have caught a significant research interest in the past few years. Owing to their attractive features including high mobility, low cost and deployment flexibility, UAVs have been used in various applications to enhance wireless communication and spectral efficiency \cite{GGr,Prabc}.

The integration of UAVs in wireless networks is studied under two research directions. The first direction deploys UAVs as moving aerial access points (APs), or moving relays, where UAVs can dynamically position themselves to boost the network's coverage, spectral efficiency and user quality of experience (QoE) \cite{SB}. The second direction considers the UAV as an aerial user-end (UE), where studies focus on the services which wireless networks can bring to the UAVs \cite{Andrea}. In both cases, the UAV must receive information from the terrestrial network. To this end, methods to enhance UAV communication in cellular networks were investigated, including massive multiple-input multiple-output (MIMO) \cite{GGr,Prabc, HYi}.

In parallel to the research in UAV communication, there is an increasing interest in transitioning from cellular to cell-free (CF) networks. A CF network consists of the deployment of large number of distributed APs that coherently serve all users with no cell boundaries, therefore fully exploiting macrodiversity and offering high probability of coverage \cite{Ngo}. Recent results show that a CF network outperforms traditional cellular and small cell networks \cite{Zhang}, thus making CF network a potential candidate for future wireless networks. Future CF networks are also expected to support UAVs, which has been recently proposed in literature. For instance, in \cite{Dandrea}, authors have investigated the perfomance of a CF-MIMO system supporting UAVs and GUEs. Results show that by proper power split between GUEs and UAVs at the AP, a CF-MIMO system may provide superior UAV communication performance over a traditional cellular MIMO system.

Despite the promising benefits, wireless communication with UAVs face challenges in terrestrial networks, since the terrestrial networks are typically designed to serve GUEs. One design parameter is the antenna down-tilt at the APs, wherein AP antennas are mechanically down-tilted to maximize the GUEs' connectivity \cite{Ma}. Under this design, the main lobes of the radiation pattern from the APs are pointing to the ground to optimize coverage for the GUEs, while the UAVs flying above the APs are only supported through the side-lobes. This affects the AP-UAV link, resulting in lower received signal power and degraded achievable rate at the UAV as shown in recent 3GPP studies \cite{Foto}. One straightforward solution is to adjust the antenna tilt and/or beam direction to support both UAVs and GUEs. However, UAVs may not be served with the required data rates due to their elevation and GUEs will suffer from performance degradation. Alternatively, one can deploy more APs to meet UAV's achievable rate requirements, however, this may be undesirable due to the additional power consumption and deployment cost. Therefore, the current terrestrial networks face a challenge in achieving desired performance at UAVs and GUEs simultaneously. This motivates further research to enhance UAV signal reception without degrading GUE's performance.





Recently, reconfigurable intelligent surfaces (RIS) have attracted a lot of attention in wireless commmunication due to its ability in reshaping the signal propagation environment so as to improve the system performance. An RIS is comprised of many passive reflecting elements, that can be digitally controlled to induce different phase-shifts on an incident signal, thus enhancing signal propagation and improving achievable rates \cite{Gong}. Most recently, RISs were considered in improving UAV communication in several applications \cite{cai,Ma}, where authors have shown that deploying an RIS can significantly enhance the received signal strength at the UAV while potentially improving the system's coverage. Though the above works on RIS-assisted UAV communication provide notable results, they strictly consider UAV based-communication. In particular, the setting where an RIS support UAVs and GUEs simultaneously has not yet been investigated in literature.

\vspace{-0.016 in}
Motivated by this consideration, we consider a scenario where an RIS is employed to support UAV communication while potentially improving the downlink (DL) achievable rate at the GUEs. Specifically, as an initial setup, we consider a CF-MIMO system with APs equipped with down-tilted antennas serving multiple GUEs and a single UAV. The main contributions of this work include the following: (i) we propose a transmission scheme that employs an RIS in a CF-MIMO system to support UAV communication without degrading GUE DL performance, but rather improving it; (ii) under conjugate beamforming (CB) at the APs, we express the DL signal-to-interference-plus-noise ratio (SINR) for the UAV and GUE; (iii) we formulate the resulting max-min rate optimization problem, and due to the non-convexity of the problem, we adopt a simplified suboptimal solution based on proportional power allocation (PPA) \cite{Ngo,Dandrea}; and (iv) we compare the DL achievable rate of our proposed system with that of two benchmark systems: one without an RIS to asses the improvement brought in by the RIS in this system, and one without a UAV and RIS to asses the impact of connecting a UAV through an RIS on the GUE performance. Interestingly, results show that under an RIS configured to serve the UAV and a proper selection of power split factor at the APs, the proposed system can improve the performance of both the UAV and GUEs, showing that the deployment of an RIS enables supporting a UAV without changing the antenna tilt at the AP and without any negative impact on GUEs but rather with a positive impact. The conclusions drawn from this work serve as preliminary insights on the capability of RISs in enhancing UAV communication in wireless networks while still benefiting the GUEs.

\vspace{-0.016 in}
In the next section, we present the system model, whose performance is analyzed later.
\begin{figure}[!t]
\centering
\begin{tikzpicture}[scale=.8]
\node (c) at (0,0,0) {};
\draw[->] (c.center) to ($(c.center)+(0,0,5)$);
\draw[->] (c.center) to ($(c.center)+(5,0,0)$);
\draw[->] (c.center) to ($(c.center)+(0,3,0)$);
\node at ($(c.center)+(0,0,5.5)$) {$x$};
\node at ($(c.center)+(5.3,0,0)$) {$y$};
\node at ($(c.center)+(0,3.2,0)$) {$z$};
\draw[fill=gray!50!white] (0,.5,2) to (0,.5,4) to (0,1.5,4) to (0,1.5,2) to (0,.5,2);
\node (r) at (0,1,3) {};
\node at (0,2,3) {\footnotesize RIS};
\node (ap1) at (3.5,0,3) {\includegraphics[scale=.3]{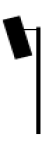}};
\node at ($(ap1.south)-(0,.1,0)$) {\footnotesize AP $m$};
\node at (5.5,0,2) {\includegraphics[scale=.3]{ap}};
\node at (5,0,3) {\includegraphics[scale=.3]{ap}};
\node (uav) at (2.5,2,0) {\includegraphics[scale=.3]{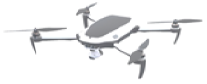}};
\node at ($(uav.north)+(0,.3,0)$) {\footnotesize UAV};
\node (ue1) at (2,0,4) {\includegraphics[scale=.3]{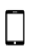}};
\node at ($(ue1.south)-(0,.1,0)$) {\footnotesize GUE $k$};
\node at (2,0,1) {\includegraphics[scale=.3]{ue}};
\node at (4.3,0,3.5) {\includegraphics[scale=.3]{ue}};
\draw[->] (ap1) to node[sloped, anchor=center,below] {\footnotesize $h_{m,k}$} (ue1);
\draw[->] (ap1) to node[sloped, anchor=center,above] {\footnotesize $\mathbf{h}_{m,\rm ris}$} (r);
\draw[->] (r) to node[sloped, anchor=center,below,fill=white] {\footnotesize $\mathbf{h}_{{\rm ris},k}$} (ue1);
\draw[->,dashed] (ap1) to node[sloped, anchor=center,above] {\footnotesize $h_{m,0}$} (uav);
\draw[->] (r) to node[sloped, anchor=center,above,fill=white] {\footnotesize $\mathbf{h}_{{\rm ris},0}$} (uav);
\end{tikzpicture}
\caption{Sketch of an RIS assisted CF-MIMO system serving GUEs/UAV.}
\label{Fig:ref}
\end{figure}
\section{System Model}
We consider a system consisting of $M$ single-antenna APs communicating with $U$ single-antenna GUEs and a single-antenna UAV (see Fig.1). We assume that the APs, GUEs, and UAV takes on a position that is uniformly distributed in a $D \times D$ $\text{m}^{2}$ geographical area and at fixed heights \cite{Ngo}. The DL transmission between the APs and UAV/GUEs is aided by an RIS that is equipped with $N$ passive reflecting elements and installed in the line-of-sight (LoS) of the APs. The RIS is mounted on a facade of an elevated building at a fixed position along one horizontal axis \cite{weii, LI}. In the following, we define the channels of this system and provide the DL achievable rates at the UAV and GUEs, which are then used as a performance metric in our system analysis.

\subsection{Communication Model}
To describe the transmission model of an RIS assisted CF-MIMO system supporting both GUEs and a UAV, we first define the channels as depicted in Fig.\ref{Fig:ref}. Let $K$ be the total number of users including the UAV, i.e., $K=U+1$, and let $k \in \{0, \ldots, U\}$ be the user indices, with $k=0$ refering to the UAV and $k = 1,\ldots,U$ refering to the $U$ GUEs. As shown in Fig.\ref{Fig:ref}, the channel between AP $m$ and user $k$ is denoted by  $h_{m,k}$, $k=0,\ldots,U$, the LoS channel between AP $m$ and the RIS is denoted by $\mathbf{h}_{m,\rm ris} \in \mathbb{C}^{N \times 1}$, and the channel between the RIS and user $k$ is denoted by $\mathbf{h}_{\rm{ris},k}\in \mathbb{C}^{N \times 1}$, $k = 0,\ldots,U$. Additionally, the response of the RIS is captured in $\boldsymbol{\Theta}=\text{diag}(\mathbf{v})\in \mathbb{C}^{N\times N}$, where $\mathbf{v}=[\zeta_1 e^{j\vartheta_{1}}, \zeta_2 e^{j\vartheta_{2}},\ldots, \zeta_N e^{j\vartheta_{N}}]^T\in \mathbb{C}^{N\times 1}$ is the RIS reflect beamforming vector, with $\vartheta_{n}\in[0,2\pi] $ being the induced phase-shift and $\zeta_{n} \in[0,1] $ being the given amplitude reflection coefficient of element $n$. To achieve the largest array gain, we assume $\zeta_{n}=1$, $\forall n=1,\ldots, N$ \cite{Deepak}. We denote components of $\mathbf{v}$ by $v_n=\exp(j\vartheta_n)$  in the remainder of this paper. 

Following the above definitions, we express the channel vector between the $M$ APs and user $k$ as
\begin{align}\label{gkk}
\mathbf{g}_{k}=\mathbf{h}_{k}+\mathbf{H}_{\rm ris}\boldsymbol{\Theta} \mathbf{h}_{\rm{ris},k},
\end{align}
where $\mathbf{h}_{k}=[h_{1,k},\ldots, h_{M,k}]^{T} \in \mathbb{C}^{M \times 1}$, and $\mathbf{H}_{\rm{ris}}=[\mathbf{h}_{1,\rm{ris}},\dots,\mathbf{h}_{M,\rm{ris}}]^{T} \in \mathbb{C}^{M \times N}$. Each entry of the channel vector $\mathbf{g}_{k}=[g_{1,k}, \ldots, g_{M,k}] \in \mathbb{C}^{M \times 1}$ denotes the channel between AP $m$ and user $k$ where
\begin{align}\label{ch1}
g_{m,k}=h_{m,k}+\mathbf{h}_{m,\text{ris}}^{T}\boldsymbol{\Theta} \mathbf{h}_{\text{ris},k}.
\end{align}
During the downlink transmission, AP $m$ employs a Gaussian codebook and linear precoding technique to transmit the data signal $x_{m}$. Using the channel in \eqref{ch1}, the downlink received signal at the UAV and GUEs is given by
\begin{align}\label{ch22}
y_{k}&=\sum_{m=1}^{M}g_{m,k}x_{m}+n_{k},
\end{align}
where $n_{k} \sim \mathcal{CN}(0,\sigma_{s}^{2})$ represents additive white Gaussian noise at the $k$'th user, and $\sigma_{s}^{2}$ is the noise power. Next, we describe the details of the downlink transmission scheme.

\subsection{Downlink Transmission}
In the downlink transmission, each AP performs beamforming using the available channel state information (CSI) and transmits a signal to all $K$ users. Let $s_{k} \sim \mathcal{CN}(0,1)$, $k= 0,\ldots,U$, where $\mathbb{E}[|s_{k}|^{2}]=1$, be the symbol intended to user $k$. Then the transmit signal from AP $m$ can be expressed as follows 
\begin{align}\label{ch33}
x_{m}=\sum_{k=0}^{U}\eta_{m,k}^{1/2}{w}_{m,k}s_{k},
\end{align}
where ${w}_{m,k}$ denote the precoder at AP $m$ assigned to user $k$, and $\eta_{m,k}$, $m=1 \ldots M$, $k=0,\ldots,U$, is power control coefficient. 
Letting $p_d$ denote the total average power available at any AP, the transmit signal power must satisfy the constraint
\begin{align}\label{pow}
\mathbb{E}[|x_{m}|^{2}]=\sum_{k=0}^{U}\eta_{m,k}\gamma_{m,k}\leq p_{d},
\end{align}
where $\gamma_{m,k}=\mathbb{E}[|{w}_{m,k}|^{2}]$ \cite{Ngo}. Note that although it seems that the UAV is treated just like an additional user, it has special properties including its channel, pathloss, elevation etc. \cite{Dandrea}, compared to the GUEs. 

The received downlink signal at user $k$, for $k=0,\ldots,U$, provided in \eqref{ch22} can be rewritten as
\begin{align}\label{rate1}
y_{k}=\sum_{m=1}^{M}\sum_{k'=0}^{U}\eta_{m,k'}^{1/2}g_{m,k}w_{m,k'}s_{k'}+ n_{k},
\end{align}
where $s_{k}$ is to be decoded from $y_{k}$ while treating the remaining $s_{k'}$, $k'\neq k$ as noise.

Using the above notations, the downlink achievable rate of user $k$ is $R_{k}=\log_{2}(1+\text{SINR}_{k})$ where $\text{SINR}_{k}$ is the signal-to-interference-plus-noise ratio at user $k$ and is given by
\begin{align}
\label{sinr}
\text{SINR}_{k}=\frac{|\sum_{m=1}^{M}\eta_{m,k}^{1/2} g_{m,k}w_{m,k}|^{2}}{\sum_{k'\neq k}|\sum_{m}^{M}\eta_{m,k'}^{1/2}g_{m,k}w_{m,k'}|^{2}+\sigma_{s}^{2}}.
\end{align}
\vspace{0.1 in}

In this work, we investigate the downlink performance of an RIS assisted CF-MIMO system supporting both GUE and UAV in terms of the achievable rate $R_{k}$. In the following section, we define the transmit precoding at the APs, RIS configuration, and power allocation adopted during the downlink transmission.





\section{Beamforming and power allocation}\label{RIS_conf}

We now provide the beamforming techniques utilized at the APs and RIS during the downlink transmission. As far as the APs are concerned, we implement CB technique, also known as maximum-ratio (MR) processing, which is a computationally simple linear precoding scheme that can be implemented in a distributed manner, i.e., locally at the APs.\footnote{Other linear processing techniques (e.g. zero-forcing) may improve the system performance, but they require more backhaul than CB. Implementing both precoding schemes and studying the tradeoff between complexity and system performance for our considered system is of interest and is left for future work.} Since the antennas are down-tilted, we choose to optimize the RIS in favor of the UAV to improve its performance. To enhance the UAV communication, we configure the RIS beamforming matrix such that the reflected signal from the RIS to the UAV add constructively with the ones recieved from the APs. The following subsections provide further details on the above adopted schemes. Furthermore, we formulate a max-min rate maximization problem at the APs and describe the adopted PPA scheme for power allocation.




\subsection{Conjugate beamforming and RIS configuration}
Below we describe the DL transmission under CB precoding at the APs and the RIS beamforming designed to serve the UAV.


\subsubsection{Transmit precoding} 
In DL transmission with CB, the signal trasmitted from the $m$th AP in \eqref{ch33} is given by
\begin{align}\label{ch34}
x_{m}=\sum_{k=0}^{U}\eta_{m,k}^{1/2}{g}_{m,k}^{*}s_{k},
\end{align}
such that the precoder $w_{m,k}$ in \eqref{ch33} is the conjugate of the true channel. In this work, we assume that the APs have the true channel knowledge, hence channel estimation and training overhead are ignored.\footnote{Channel estimation can be performed by finding the minimum mean square error (MMSE) estimates of the aggregated channel using the direct estimation (DE) protocol proposed in \cite{bayan}. The advantage of this protocol is the reduced training overhead. Performance analysis of our proposed system under imperfect CSI and the derivation of the a closed-form expression of the DL SINR is left as a future extension of this work.} Under CB precoding, the received DL signal at user $k$ given in \eqref{rate1} can be expressed as
\begin{align}\label{ch3334}
y_{k}=\sum_{m=1}^{M}\eta_{m,k}^{1/2}g_{m,k}g_{m,k}^{*}s_{k'}+\sum_{k'\neq k} \sum_{m=1}^{M}\eta_{m,k}^{1/2}g_{m,k}g_{m,k'}^{*}s_{k'}+n_{k},
\end{align}
where the first term is the desired signal, and the second and third terms are treated as the effective noise. Then, the $\text{SINR}_{k}$ expression in \eqref{sinr} under CB can be expressed by setting $w_{m,k}=g_{m,k}^{*}$. We use $\text{SINR}_{k}$ to study the DL performance of our considered system and investigate the impact of system parameters such as number of RIS reflecting elements and the power split factor at the APs on the DL achievable rate at the UAV and GUEs.



\subsubsection{RIS configuration} To support the UAV-communication, we configure the RIS to direct the reflected signal towards the UAV. For our considered single UAV system, the vector $\mathbf{v}=[\exp(j\vartheta_{1}), \exp(j\vartheta_{2}),\ldots, \exp(j\vartheta_{N})]^T$ is designed so as the reflected signals from the RIS add constructively with the ones received directly from the APs. Let $\mathbf{R}=\mathbf{H}_{\text{ris}}^{T}\text{diag}(\mathbf{h}_{\text{ris},0})\in \mathbb{C}^{M \times N}$. Therefore, we can rewrite the channel between the APs and the UAV in \eqref{ch1} as $\mathbf{g}_{0}=\mathbf{h}_{0}+\mathbf{R}\mathbf{v}$. The received signal power at the UAV can then expressed as
\begin{align}\label{eqref1}
P_{\mathcal{U}}=|(\mathbf{R}\mathbf{v}+\mathbf{h}_{0})^{T}\mathbf{w}_{0}|^{2}
\end{align}
where $\mathbf{w}_{0}=\mathbf{g}_{0}^{*}$ is the precoder vector. A desirable choice of $\mathbf{v}$ would be one that maximizes \eqref{eqref1}. Authors in \cite{Deepak} proposed a near-optimal RIS configuration, which helps to maximize the recieved power for a single user by aligning the reflected signal from the RIS with the direct signal from the transmitter. Accordingly, by adopting their approach in our single UAV system, the RIS beamforming vector $\mathbf{v}$ can be defined as
\begin{align}
\mathbf{v}^{*} =\text{exp}(-j\angle{\mathbf{R}^{T}}\mathbf{h}_{0}^{*}),
\end{align}
and $\angle{\mathbf{x}}$ returns the vector of phases of $\mathbf{x}$. We consider the above RIS configuration as one possible solution to maximize \eqref{eqref1}. Note that in this work, we consider CB at the APs and optimize the RIS reflect beamforming vector for a given transmit precoder. As seen later in Sec. IV, this (sub-optimal) approach proves to provide promising gains in DL performance for both UAVs and GUEs under an appropriate transmit power splitting at the APs . 

In the following section, we formulate a power allocation optimization problem under the considered RIS configuration and CB precoding and adopt PPA as a solution to the problem.




\subsection{Power allocation} \label{secc11}
We formulate the max-min rate maximization problem criterion to optimize the power allocation, similar to other works in literature \cite{Andrea, Ngo}. Since each AP has a transmission power limit in practical systems, we use the per AP power constraints as in \eqref{pow}. Let $\boldsymbol{\eta}=[\eta_{m,k}:m=1,\ldots,M, k=0,\ldots,U]$ be a positive vector containing the power coefficients. The max-min rate maximization problem can be expressed as
\begin{alignat}{2} \textit{(P1)} \hspace{.1 in}
&\!\max_{\boldsymbol{\eta}} \min_{k} \hspace{0.1 in}  \text{R}_{k}(\boldsymbol{\eta}) \label{aaf}\\
&\text{s.t.} \hspace{0.03 in} \sum_{k=0}^{U}\eta_{m,k}\gamma_{m,k}\leq p_{d}, \forall m= \{1,\ldots ,M\}\nonumber. 
\end{alignat}
\begin{table}[t!]\label{T1}
\centering
\caption{System Parameters}
\begin{tabular}{ |c|c|c| } 
\hline
 Parameter & Value \\
 \hline 
 carrier frequency & 1.9 GHz  \\ 
 \hline
 bandwidth & 20 MHz \\ 
 \hline
 $\xi_{m,k}$ & Hata-COST23 model \cite{Ngo} \\ 
  \hline
  $\rho$, $\alpha$ & -30 [dB], 2.4 \cite{WuQ} \\ 
    \hline
 $K_{t,k}$, $t\in \{ris, m\}$ & $13-0.03d_{t,k}$ \cite{Oz2} \\ 
 \hline
 $\sigma_{s}^{2}$ & -62 dBm \\ 
 \hline
 ${p}_{d}$ &  1 W  \\ 
  \hline  
\end{tabular}
\label{T1}
\end{table}

The problem in $(P1)$ is a nonconvex optimization problem and is difficult to optimize using conventional methods. For the purpose of this work, we simplify the problem by adopting a baseline power allocation strategy known as proportional power allocation (PPA), where the $m$-th AP shares the transmit power $p_{d}$ in a way that is proportional to the channel strengths. By letting $P^{DL}_{m,k}=\eta_{m,k}\gamma_{m,k}$ denote the power transmitted by the $m$-th AP to the $k$-th user, the PPA policy proposed in \cite{Dandrea} under our system setup becomes, 
\vspace{0.05 in}
\[
  P_{m,k}^{DL} =
  \begin{cases}
                                   (1-\kappa)p_{d}\frac{\gamma_{m,k}}{\sum_{j=1}^{U}\gamma_{m,j}}& \text{if $k\in\{1,...,U\}$} \\
                                   \kappa p_{d}& \text{if $k=0$} 
  \end{cases}
\]

\vspace{0.05in}
\noindent where $\kappa$ is the fraction of the power allocated by the AP to serve the UAV.  Note that since the UAV benefits from the RIS, which is tuned in its favor, the $\kappa$ parameter helps to provide a further degree of freedon in terms of dividing the power resource to guarantee GUE's performance. 

\section{Simulation}\label{Sec:Sim}

In this section, we numerically evaluate the UAV and GUEs DL performance in an RIS-assisted CF-MIMO system, and compare it to a baseline CF-MIMO system. We consider $M=20$ APs, $K=5$ users, with $U=4$ GUEs and a single UAV, uniformly distributed in a $40 \times 40 \text{m}^{2}$ area. To simulate an RIS-assisted CF-MIMO system, we additionally deploy an RIS that is equipped with $N$ reflecting elements to assist in the communication between the APs and UAV/GUEs. In the following, we describe the system setup with channel models, then present the numerical results.

\subsection{Simulation Parameters}

We adopt a cartesian coordinate system where all communication nodes locations are defined by two horizontal coordinates and one vertical coordinate. The GUEs and UAV horizontal coordinates are denoted by $\mathbf{q}_{k}=[q_{k}^{(x)}, q_{k}^{(y)}]$, for $k=0,\ldots U$, while the AP and RIS horizontal coordinates are denoted by $\mathbf{q}_{\rm AP,m}=[q_{\rm AP,m}^{(x)}, q_{\rm AP,m}^{(y)}]$ for $m=1,\ldots ,M$  and $\mathbf{q}_{\rm ris}=[q_{\rm ris}^{(x)}, 0]$, respectively. The vertical coordinates are denoted by $H_{k}$, $H_{\rm AP,m}$ and $H_{\rm ris}$, for user $k$ (GUEs and UAV), AP and RIS, respectively. While the horizontal coordinates are uniformly distributed, the AP, RIS and GUE heights are fixed to $H_{\rm AP,m}= 15$m, $H_{\rm ris}=12$m and $H_{k}=1.65$m, $k=1, \dots, U$, respectively. Since UAVs typically fly at an altitude of $1.5-300$m based on 3GPP standard \cite{GGr}, we fix the UAV height to $H_{0}= 100$m in our simulation results, unless otherwise noted. Therefore, we can define the Euclidean distance between the nodes as in \cite{Andrea} with $d_{m,k}$, $d_{\text{ris},k}$ and $d_{m,\text{ris}}$ denoting the distance between AP $m$ and user $k$, RIS and user $k$ and AP $m$ and RIS, respectively, for $k=0,\ldots ,U$ and $m=1,\ldots, M$. 

The channel between AP $m$ and GUE, i.e., $k=1,\ldots,U$, is denoted as
\begin{align}\label{ch8}
h_{m,k}=\sqrt{\zeta_{m,k}} (\sqrt{\frac{K_{m,k}}{K_{m,k}+1}}\bar{h}_{m,k}+\sqrt{\frac{1}{K_{m,k}+1}}\tilde{h}_{m,k}),
\end{align}
where $\zeta_{m,k}=10^{\frac{A(\theta_{m,k})}{10}}\xi_{m,k}$, $A(\theta_{m,k})$ [dB] is the antenna downtilt model based on the 3GPP TR 36.814 \cite{3GPP}, $\xi_{m,k}$ is the channel attenuation factor, $\theta_{m,k}$ is the angle of arrival (AoA) between AP $m$ and GUE, $\bar{h}_{m,k}=e^{-j\frac{2\pi d_{m,k}}{\lambda}}$ and $\tilde{h}_{m,k}\sim \mathcal{CN}(0,1)$ denote the deterministic LoS component and the small-scale fading component, respectively, $K_{m,k}$ is the Rician factor, and $\lambda$ is the carrier wavelength. Table \ref{T1} provides the system parameter definitions for the reader's reference. Similarly, the channel between AP $m$ and UAV can be expressed as \eqref{ch8} with $k=0$ and $\xi_{m,0}=\rho d_{m,0}^{-\alpha}$, where $\rho$ is the pathloss of reference distance $1$m \cite{WuQ} and $\alpha$ is the pathloss exponent. The LoS channel between the AP $m$ and RIS is expressed as $\mathbf{h}_{m,\rm{ris}}=\sqrt{10^{\frac{A(\theta_{m,\text{ris}})}{10}}\rho d_{m,\text{ris}}^{-\alpha}}\mathbf{\bar{h}}_{m,\text{ris}} \in \mathbb{C}^{N\times 1}$, where $\theta_{m,\text{ris}}$ is the AoA between AP $m$ and the centre of the RIS and $\mathbf{\bar{h}}_{m,\text{ris}}$ is the array response vector as defined in \cite{Hua}. Finally, the channel vector between the RIS and GUE/UAV, for $k=0, \ldots, U$, can be expressed as $\mathbf{h}_{\rm{ris},k}=\sqrt{\rho d_{\rm{ris},k}^{-\alpha}}(\sqrt{\frac{K_{\rm{ris},k}}{K_{\rm{ris},k}+1}}\bar{\mathbf{h}}_{\rm{ris},k}+\sqrt{\frac{1}{K_{\rm{ris},k}+1}}\tilde{\mathbf{h}}_{\rm{ris},k})$, where $\bar{\mathbf{h}}_{\rm{ris},k} \in \mathbb{C}^{N\times 1} $ is the array response vector and $\tilde{\mathbf{h}}_{\rm{ris},k} \sim \mathcal{CN}(0,I_{N})$ is the small-scale fading component, respectively. Assuming RIS and UAV are elevated, we define the channel between RIS and UAV, i.e., $\mathbf{h}_{\rm{ris},0}$, as a LoS channel by approaching $K_{\rm{ris},0}$ to infinity \cite{LI,weii}.

Given the above definitions, we can now define the system parameter $\gamma_{m,k}$, $k=0, \dots, U$, used in the PPA scheme. Recall that $\gamma_{m,k}=\mathbb{E}[|{w}_{m,k}|^{2}]$. Using the definitions of UAV and GUE precoder, i.e., $w_{m,k}=g_{m,k}^{*}$ for $k=0,\ldots, U$, we can express $\gamma_{m,k}=|\mu_{m,k}|^{2}+\sigma^{2}_{m,k}$, where $\mu_{m,k}=\sqrt{\frac{K_{\rm{ris},k}\beta_{\rm{ris},k}^{2}}{K_{\rm{ris},k}+1}}\mathbf{h}_{m,\rm{ris}}^{T}\boldsymbol{\Theta}\bar{\mathbf{h}}_{\rm{ris},k}+\sqrt{\frac{K_{m,k}\beta_{m,k}^{2}}{K_{m,k}+1}}\bar{h}_{m,k}$, $\beta_{m,k}=\sqrt{\zeta_{m,k}}$, $\beta_{\text{ris},k}=\sqrt{\rho d_{\rm{ris},k}^{-\alpha}}$, and $\sigma^{2}_{m,k}=\frac{\beta_{m,k}^{2}}{K_{m,k}+1}+\frac{\beta_{\rm{ris},k}^{2}}{K_{\rm{ris},k}+1}\mathbf{h}_{m,\rm{ris}}^{T}\boldsymbol{\Theta}\boldsymbol{\Theta}^{H}(\mathbf{h}_{m,\rm{ris}}^{H})^{T}$ for $k=1, \ldots, U$.
\vspace{0.01 in}

In the following section, we provide numerical results for demonstrating the DL performance of our proposed system under the defined system parameters. 



\subsection{Numerical Results}
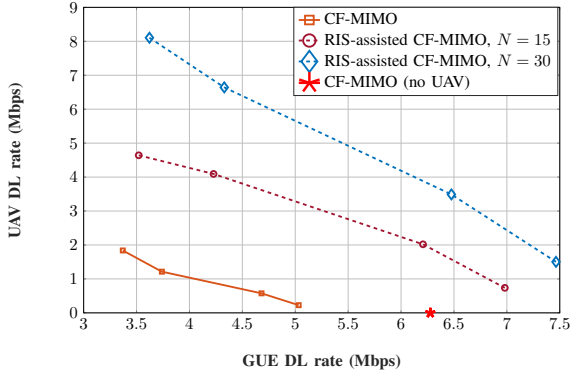
\begin{figure}[t!]
\tikzset{every picture/.style={scale=.47}, every node/.style={scale=1.34}}
%
%
\definecolor{mycolor1}{rgb}{0.85000,0.32500,0.09800}%
\definecolor{mycolor2}{rgb}{0.63500,0.07800,0.18400}%
\definecolor{mycolor3}{rgb}{0.00000,0.44700,0.74100}%
\begin{tikzpicture}

\begin{axis}[%
width=11.281in,
height=7.244in,
at={(1.645in,1.18in)},
scale only axis,
xmin=3,
xmax=7.5,
xlabel style={font=\color{white!15!black}},
ylabel style={font=\color{white!15!black}},
xlabel={GUE DL rate (Mbps)},
ymin=0,
ymax=9,
xlabel style={font=\bfseries\color{white!15!black}, at={(ticklabel cs:0.5)}},
ylabel style={font=\bfseries\color{white!15!black}, at={(ticklabel cs:0.5)}},
ylabel={UAV DL rate (Mbps)},
axis background/.style={fill=white},
xtick={3, 3.5, 4, 4.5, 5, 5.5, 6, 6.5, 7, 7.5},
ytick={0, 1, 2, 3, 4,5, 6, 7, 8, 9},
xmajorgrids,
ymajorgrids,
legend style={at={(2.12,2.13)}, legend cell align=left, align=left, draw=white!15!black}
]
\addplot [color=mycolor1, line width=1.5pt, mark size=3.5pt, mark=square, mark options={solid, mycolor1}]
  table[row sep=crcr]{%
3.36993 1.83792\\
3.73937	1.21344\\
4.6812	0.575288\\
5.032411	0.22649\\
};
\addlegendentry{CF-MIMO}

\addplot [color=mycolor2, dashed, line width=1.5pt, mark size=5.0pt, mark=o, mark options={solid, mycolor2}]
  table[row sep=crcr]{%
3.52069	4.64201\\
4.22787	4.092625\\
6.20751	2.0195\\
6.98061	0.737\\
};
\addlegendentry{RIS-assisted CF-MIMO, $N=15$}

\addplot [color=mycolor3, dashed, line width=1.5pt, mark size=8.6pt, mark=diamond, mark options={solid, mycolor3}]
  table[row sep=crcr]{%
3.622	8.10495\\
4.329	6.64244\\
6.47571	3.489\\
7.4643	1.50263\\
};
\addlegendentry{RIS-assisted CF-MIMO, $N=30$}

\addplot [color=red, line width=2.0pt, mark size=8.6pt, mark=star, mark options={solid, red}]
  table[row sep=crcr]{%
6.2766375	0\\
6.2766375	0\\
6.2766375	0\\
6.2766375	0\\
};
\addlegendentry{CF-MIMO (no UAV)}

\end{axis}

\begin{axis}[%
width=14.556in,
height=8.889in,
at={(0in,0in)},
scale only axis,
xmin=0,
xmax=1,
ymin=0,
ymax=1,
axis line style={draw=none},
ticks=none,
axis x line*=bottom,
axis y line*=left,
legend style={legend cell align=left, align=left, draw=white!15!black}
]
\end{axis}
\end{tikzpicture}%
\caption{DL achievable per-user rate region achieved by the UAV and GUE over $\kappa=[0.02, 0.05, 0.1, 0.15]$ and under $95\%$ probability of coverage.}
\label{fig1}
\end{figure}
Fig. \ref{fig1} plots the DL achievable per-user rate regions obtained for the GUE and UAV under varying values of $\kappa$ and for three different systems of interest. In particular, in Fig. \ref{fig1} we study the DL achievable per-user rates for a standard CF-MIMO system and RIS-assisted CF-MIMO system serving both GUEs and a UAV, as well as a standard CF-MIMO system serving only GUEs without a UAV. We plot the DL achievable per-user rate regions achieved by the UAV and GUE over $\kappa=[0.02,0.05, 0.1, 0.15]$ under a $95\%$ probability of user-coverage. Here, without loss of generality, we define the per-user rate as the rate achieved by GUE user 1 ($k=1$) and the UAV ($k=0$) \cite{NN2}.  

To further elaborate on the results shown in Fig. \ref{fig1}, we first look at the solid line (with square markers), which depicts the DL achievable rate region for a standard CF-MIMO system. We see that for low values of $\kappa$, the GUE achieves significantly higher $95\%$-likely DL per-user rate compared to that of the UAV. As $\kappa$ increases, i.e., allocating more transmit power to UAV than GUE, the UAV DL rate improves while the GUE DL performance degrades. This phenomenon is expected and shows the tradeoff between the GUEs and the UAV DL performance in CF-MIMO system serving both ground and aerial users. Second, we look at RIS-assisted CF-MIMO system depicted by the dotted curves (with circle and diamond markers), where RIS is equipped with $N=15$ and $N=30$, respectively. The following observations can be drawn from the curves when comparing them to that of the standard CF-MIMO. First, it can be seen that while the DL achievable rate of both the GUE and UAV improves under RIS deployment for low values of $\kappa$, the DL achievable rate of the UAV improves more compared to GUE for high values of $\kappa$. That is mainly due to the presence of the RIS that is configured to enhance the communication to the UAV. Second, it is interesting to observe that for increasing values of $\kappa$, the UAV DL rate improves more that $3$ folds reaching an almost equivalent DL rate as that of the GUE. Moreover, as $N$ increases, both UAV and GUE benefit from the RIS array gains. Finally, we consider a CF-MIMO system  without a UAV (depicted by a star) and compare the GUE DL achievable rate to that of the RIS-assisted CF-MIMO system serving both UAV and GUEs. We see that for low values $\kappa$, the DL performance of the GUE in an RIS-assisted CF-MIMO system (with $N=30$) can achieve better than that served in a CF-MIMO system without a UAV, while at the same time we see that the UAV benefit from the RIS gains. The main takeaway from Fig. \ref{fig1} is that under proper selection of $\kappa$ and number of reflecting elements $N$, an RIS-assisted CF-MIMO system can significantly improve the UAV communication, while at the same time improving the GUE DL rate. 
\begin{figure}[t!]
\tikzset{every picture/.style={scale=.47}, every node/.style={scale=1.34}}
\input{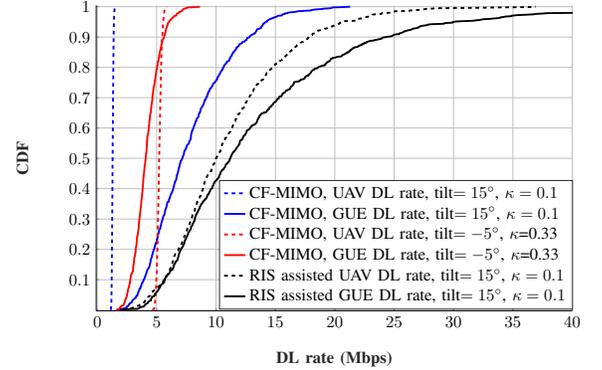}
\caption{CDF of the DL rates in RIS-assisted CF-MIMO and CF-MIMO systems under varying $\kappa$ and AP antenna tilt. Here $N=20$.}
\label{fig2}
\end{figure}
In Fig. \ref{fig2}, we study the effect of $\kappa$ and AP antenna down-tilt on the UAV and GUE DL per-user rates for a standard CF-MIMO system (no RIS). Then, we examine the impact of deploying an RIS to compensate the effect of AP down-tilt and power allocation on the UAV DL performance. Namely, we first consider $\kappa=0.1$ and an AP antenna down-tilt of $15^{\circ}$ based on 3GPP model \cite{GGr}, and plot the CDF of the DL rates of the UAV (dotted-blue curve) and per-user GUE (solid blue curve). The CDF curves are plotted over uniformly random UAV and GUEs positions. To serve the UAV better, we then consider tilting the AP antennas upward towards the UAV such that the antenna tilt is $-5^{\circ}$ and allocate more power at the APs to serve the UAV ($\kappa=0.33$). By examinig the CDF curves in Fig. \ref{fig2}, we see that while UAV benefits from adjusted system settings, the GUE DL rate largely degrades. However, by deploying an RIS equipped with reflecting elements configured to serve the UAV and by keeping $\kappa=0.1$ and keeping the APs antenna down-tilt at $15^{\circ}$, we can achieve improvement in the UAV as well as the GUE DL rate without having to change the AP anetnna tilt angle that is set by the operator.
\begin{figure}[t!]
\tikzset{every picture/.style={scale=.45}, every node/.style={scale=1.34}}
%
%
\definecolor{mycolor1}{rgb}{0.00000,0.44700,0.74100}%
\definecolor{mycolor2}{rgb}{0.85000,0.32500,0.09800}%
\definecolor{mycolor3}{rgb}{0.92900,0.69400,0.12500}%
\begin{tikzpicture}

\begin{axis}[%
width=11.281in,
height=7.244in,
at={(1.526in,0.849in)},
scale only axis,
xmin=20,
xmax=60,
xlabel style={font=\color{white!15!black}},
xlabel={$\mathbf{N}$ Reflecting Elements},
ymin=2,
ymax=18,
ylabel style={font=\color{white!15!black}},
xlabel style={font=\bfseries\color{white!15!black}, at={(ticklabel cs:0.5)}},
ylabel style={font=\bfseries\color{white!15!black}, at={(ticklabel cs:0.5)}},
ylabel={RIS gain [dB]},
axis background/.style={fill=white},
xmajorgrids,
ymajorgrids,
legend style={at={(0.01,1.7)}, anchor=south west, legend cell align=left, align=left, draw=white!15!black}
]

\addplot [color=mycolor2, line width=1.5pt, mark size=8.6pt, mark=diamond, mark options={solid, mycolor2}]
  table[row sep=crcr]{%
20	7.16036123190391\\
30	11.3460026316937\\
40	13.511639574141\\
50	16.1531674322473\\
60	17.6073058631447\\
};
\addlegendentry{UAV height $H_{0}=300$m}

\addplot [color=mycolor1, line width=1.5pt, mark size=3.5pt, mark=square, mark options={solid, mycolor1}]
  table[row sep=crcr]{%
20	5.64097591914794\\
30	8.33578336968008\\
40	10.8246900501538\\
50	12.7171375927624\\
60	13.9089380139815\\
};
\addlegendentry{UAV height $H_{0}=100$m}

\addplot [color=mycolor3, line width=1.5pt, mark size=5.0pt, mark=o, mark options={solid, mycolor3}]
  table[row sep=crcr]{%
20	2.53899531136697\\
30	4.44489560682835\\
40	5.7030000560815\\
50	7.2940148919945\\
60	8.00771467228575\\
};
\addlegendentry{UAV height $H_{0}=16$m}

\end{axis}

\begin{axis}[%
width=11.736in,
height=7.722in,
at={(0in,0in)},
scale only axis,
xmin=0,
xmax=1,
ymin=0,
ymax=1,
axis line style={draw=none},
ticks=none,
axis x line*=bottom,
axis y line*=left,
legend style={legend cell align=left, align=left, draw=white!15!black}
]
\end{axis}
\end{tikzpicture}%
\caption{RIS SINR gain achieved in the UAV DL SINR in an RIS- assisted CF-MIMO under varying UAV heights and number of reflecting elements $N$. Here $\kappa=0.1$.}
\label{fig4}
\end{figure}
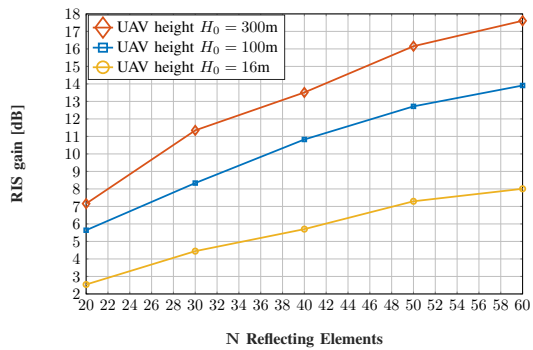

Finally, in Fig. \ref{fig4} we plot the RIS SINR gain defined as the ratio of received SINR at the UAV with and without an RIS, i.e., $\text{SINR}_{0}/\text{SINR}_{0,nonRIS}$, and use it as a metric to evaluate the performance of UAV communication under varying UAV heights and number of RIS reflecting elements in an RIS-assisted CF-MIMO system. Specifically, $\text{SINR}_{0}$ is the SINR achieved at the UAV in an RIS-assisted CF-MIMO as expressed in \eqref{sinr} and $\text{SINR}_{0,nonRIS}$ is the equivalent SINR achieved at the UAV in a standard CF-MIMO system (without an RIS) \cite{Dandrea}. Fig. \ref{fig4} shows that the RIS gain at the UAV increases as the number of RIS elements increases, for all three UAV heights. Additionally, from Fig. \ref{fig4}, we observe that as the UAV height decreases, the RIS gain drops. This is because at lower altitudes, for example $H_{0}=16$m, the UAV becomes within the main lobe of the down-tilted AP antennas, which reduces the RIS gains.

\section{Conclusion}
In this paper, an RIS-assisted CF-MIMO system supporting UAV and GUEs simultaneously was considered. We studied the potential of RIS to support UAV-communication without degrading the GUE DL performance. Under CB precoding at the APs and an RIS configured to serve the UAV, we showed that the RIS-assisted CF-MIMO system can significantly enhance the UAV DL performance while positively impacting the GUEs. We support our conclusion with numerical results and provide insight on DL performance under varying number of RIS elements, UAV heights and power allocation factor. Our results show that under a proper power split between GUEs and UAV, and number of RIS reflecting elements, the RIS-assisted CF-MIMO system can enhance the DL performance of both the UAV and GUE, while preserving the 3GPP AP antenna down-tilt. In the future extension of this work, we consider a joint transmit and reflect beamforming optimization to maximize the received desired signal power at the UAV.




\bibliographystyle{IEEEtran}
\bibliography{bibl}
\end{document}